\documentstyle[prl,aps,twocolumn,epsf]{revtex}

\begin{document}
\draft


\noindent
{\bf Comment on ``Density Functional Simulation of a Breaking Nanowire''}

\vspace{3mm}

In a recent Letter \cite{nbhj}, Nakamura et al.\ described first principles
calculations for a breaking Na nanocontact.  Their system consists of a 
periodic one-dimensional 
array of supercells, each of which contains 39 Na atoms, 
originally forming a straight, crystalline wire with a length of 6 atoms.
The system is elongated by increasing the length of the unit cell.  At each
step, the atomic configuration is relaxed to a new local equilibrium,
and the tensile force is evaluated from the change of the total energy with
elongation.  
Aside from a discontinuity of the force occuring at the transition from a
crytalline to an amorphous configuration during the early stages of elongation,
they were unable to identify any simple correlations between the force and
the number of electronic modes transmitted through the contact.
An important question is whether 
their model is realistic, i.e., whether it can be compared to experimental
results \cite{nanexp}
obtained for a single nanocontact between two macroscopic pieces
of metal.  In this Comment, we demonstrate that with such a small unit cell,
the interference effects between neighboring contacts are of the 
same size as the force oscillations in a single nanocontact. 

In order to understand how the close proximity of the nanocontacts in the
model of Ref.\ \onlinecite{nbhj} may alter the energetics of the system,
we consider a system of two identical nanocontacts in series, connecting
two macroscopic wires.  We model the 
metallic nanocontacts as constrictions in a free electron gas, with hard-wall
boundary conditions, and obtain the energetics of the system
from the electronic scattering matrix \cite{sbb}.  The scattering matrix of 
the compound system may be obtained as a geometric series in the scattering
matrices of the individual contacts (which are taken to be symmetric under
inversion, for simplicity), while 
the scattering matrix of a single contact may be evaluated using the 
adiabatic and WKB approximations \cite{sbb}, which are quite accurate for
contacts of smooth shape \cite{bszb}.
The total grand canonical potential
of the system is found to
be the sum of the contributions of the individual contacts, 
plus an interference term
\[
\Delta\Omega = -\frac{2}{\pi} \int dE f(E) \sum_\nu
\tan^{-1} \frac{R_\nu (E) \sin[2\theta_\nu(E)]}{1+R_\nu(E)\cos[2\theta_\nu(E)]},
\]
where $f(E)$ is the Fermi-Dirac distribution function, and
$R_\nu(E)$ and $\theta_\nu(E)$ are the reflection probability and 
scattering phase shift, respectively, of the $\nu$th electronic mode 
for a single nanocontact.  

The magnitude of the 
correction to the cohesive force in the supercell arrangement of Ref.\ 
\onlinecite{nbhj} arising from interference effects between neighboring
supercells is $\Delta F = - \partial [\Delta \Omega]/
\partial L_{\rm cell}$, where $L_{\rm cell}$ is the unit cell length.  
Interference between more widely separated supercells would lead to an
additional correction.  Fig.\ 1(b) shows that for
the unit cell size considered in Ref.\ \onlinecite{nbhj}
($L_{\rm cell}=17$---$31\AA=2.5$---$4.5 \lambda_F$),
the interference correction to the cohesive force is comparable to the 
force oscillations of an individual nanocontact.  
For comparison, the conductance of a single 
nanocontact and the interference correction thereof are shown in Fig.\ 1(a).
For a single contact, there is a clear correlation between the conductance
steps and the force oscillations.  However, the large interference correction
would strongly suppress any correlations between the force calculated in 
the supercell arrangement of Ref.\ \onlinecite{nbhj} and the conductance of
a single contact.  

In order to explain the correlations between cohesion and conductance 
observed experimentally in metallic nanocontacts \cite{nanexp}, it is 
essential to treat the energetics and transport of the system on an equal 
footing.  
This has been achieved in our free-electron model
\cite{sbb}.
The interference term scales as $\Delta F \sim {\cal O}(L_{\rm
cell}^{-1})$ [since $\theta_\nu(E)\propto L_{\rm cell}$], so it would be
worthwhile to perform larger-scale ``first principles'' simulations to
address this question.

J.\ B.\ acknowledges support from Swiss National Foundation 
PNR 36 ``Nanosciences'' grant \# 4036-044033. 

\vspace{3mm}

\noindent
C.~A. Stafford,$^{1,2}$
J. B\"urki,$^{2,3}$ and D.\ Baeriswyl$^2$

$^1$University of Arizona, Tucson, Arizona 85721

$^2$Universit\'{e} de Fribourg, 1700 Fribourg, Switzerland

$^3$IRRMA, EPFL, 1015 Lausanne, Switzerland

\vspace{3mm}

\noindent
Received 19 August 1999


\begin{figure}
\vspace*{-3.0cm}
\epsfxsize=9cm
\epsffile{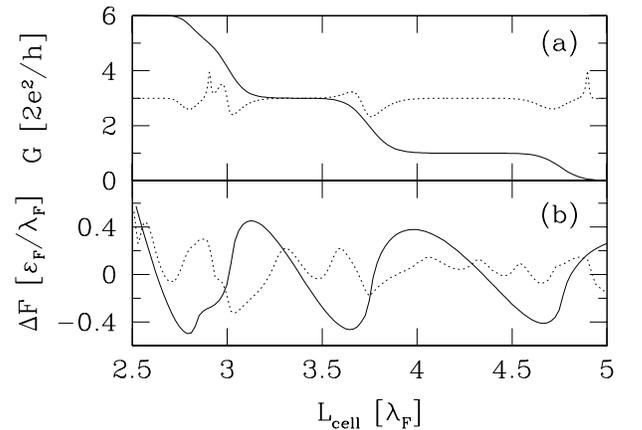}
\vspace{-3mm}
\caption{Conductance and force 
for a metallic nanocontact with dimensions 
comparable to that of Ref.\ [1]: initial radius=$\lambda_F$, 
initial length=$2.5\lambda_F$.  
(a) Conductance for a single contact (solid curve),
and interference term (dotted curve, offset by 3);
(b) force oscillations for a single nanocontact [3] (solid
curve), and 
interference term $\Delta F=-\partial [\Delta\Omega]/\partial L_{\rm
cell}$ (dotted curve). 
}
\label{fig1}
\end{figure}


\vspace{-0.5cm}

\begin{references}
\vspace{-1.5cm}

\bibitem{nbhj} A. Nakamura, M. Brandbyge, L. B. Hansen, and K. W. Jacobsen,
Phys. Rev. Lett. {\bf 82}, 1538 (1999).

\bibitem{nanexp} C. Rubio, N. Agra\"{\i}t, and S. Vieira, Phys. Rev. Lett.
{\bf 76}, 2302 (1996); A. Stalder and U. D\"urig, Appl. Phys. Lett. {\bf 68},
637 (1996).

\bibitem{sbb} C. A. Stafford, D. Baeriswyl, and J. B\"urki, Phys. Rev. Lett.
{\bf 79}, 2863 (1997).

\bibitem{bszb} J. B\"urki, C. A. Stafford, X. Zotos, and D. Baeriswyl,
Phys. Rev. B {\bf 60}, 5000 (1999).

\end{references}
\end{document}